\documentclass[aps,pre,groupedaddress,twocolumn,longbibliography,showpacs]{revtex4-1}

\usepackage{amsmath,amssymb,amsthm,graphicx,color,epstopdf}
\usepackage{natbib}
\usepackage{hyperref}
\usepackage[titletoc,title]{appendix}
\hypersetup{
	colorlinks   = true,
	citecolor    = blue,
	linkcolor = blue,
	colorlinks = true,
}

\newcommand{\dx}{\delta x}
\newcommand{\Dx}{\Delta x}
\newcommand{\dt}{\delta t}
\newcommand{\Dt}{\Delta t}
\newcommand{\dd}{\delta}

\newcommand{\ra}{\rightarrow}

\newcommand{\PP}{ \mathrm{Pr} }
\newcommand{\EE}{\mathbb{E}}
\newcommand{\up}{\uparrow}
\newcommand{\dn}{\downarrow}
\newcommand{\erf}{\text{erf}}

\begin{document}

\title{Infrequent social interaction can accelerate the spread of a persuasive idea}


\author{James Burridge}
\author{Micha\l \ Gnacik}
\affiliation{Department of Mathematics, University of Portsmouth, Lion Terrace, Portsmouth PO1 3HF, United Kingdom}



\begin{abstract}
	We study the spread of a persuasive new idea through a population of continuous-time random walkers in one dimension. The idea spreads via social gatherings involving groups of nearby walkers who act according to a biased ``majority rule'': After each gathering, the group takes on the new idea if more than a critical fraction $\frac{1-\varepsilon}{2} < \frac{1}{2}$ of them already hold it; otherwise they all reject it. The boundary of a domain where the new idea has taken hold expands as a traveling wave in the density of new idea holders. Our walkers move by L\'{e}vy motion, and we compute the wave velocity analytically as a function of the frequency of social gatherings and the exponent of the jump distribution. When this distribution is sufficiently heavy tailed, then, counter to intuition, the idea can propagate faster if social gatherings are held less frequently. When jumps are truncated, a critical gathering  frequency can emerge which maximizes propagation velocity. We explore our model by simulation, confirming our analytical results.    
	
\end{abstract}


\maketitle

\section{Introduction}

The spread of new ideas through populations is a driver of human progress and cultural change \cite{Col16}. The spread of culture (which includes ideas, opinions, language, and behavior)  involves two key processes: movement-migration and communication-copying.  In fact, the processes by which ideas spread have much in common with the processes that drive physical systems of interacting particles (thought of as \textit{agents} in the social context).  For example, social mimicking of ideas is analogous to alignment of spins in magnetic materials \cite{Hua87, Cas09, Gal91}, the migration and daily movement of individuals is statistically similar to the random motion of particles \cite{Bro06, Man00}, and the cascading spread of a new fashion or idea may be viewed as a percolation or branching process \cite{Iri11,Gle14,Gle16}. A feature of physical systems where particles tend to align---which appears to reflect a social reality---is the formation of geographical regions (domains) where one particular alignment, idea, opinion, language, \textit{etc}., is dominant to the exclusion of others. Where two domains meet, if one idea is more persuasive than the other, then the domain boundary will move so the more persuasive domain expands. Our aim is to show, analytically, how movement and interaction frequency can affect the rate at which this takes place. 

A number of statistical physics models have been introduced over the past few decades to study opinion dynamics. Among the most well known are the voter model \cite{Cli73,Cas09,Red10} where opinions evolve by copying randomly selected neighbors, and the majority rule \cite{Gal02_2,Kra03,Che05}, where groups of agents update their opinion to match the majority of the group.  Many others exist \cite{Cas09}.  Because of its simplicity, the majority rule model is easy to generalize. For example, it has been studied on networks \cite{Lam07}, with diffusing agents \cite{Gal02,Sta02}, with variable numbers of neighbors \cite{Tes04},  and on $d$-dimensional lattices \cite{Che05}, where it has strong similarities with zero temperature Ising Glauber (IG) dynamics \cite{Gla63} (the kinetic Ising model). The difference lies in the fact that under IG dynamics, spins flip one at a time, rather than in groups, in order to match the majority in their neighborhood. 

In this study we employ a biased version of majority rule dynamics to study the spread of an idea within a population of continuous-time random-walking agents on a line.  Interactions take the form of social gatherings, held between groups of nearby walkers. Within each gathering, each walker reveals whether he or she holds the new idea, after which the state of the group is decided by a biased majority rule: All walkers accept the idea provided a sufficiently large fraction already hold it. Otherwise, the idea is rejected by the entire group. Walkers retain their opinion states (accepting or rejecting the idea) between interactions, so the opinions expressed at each gathering carry information about the system at various times and locations in the past.  A real-world interpretation is that some topic, about which a new way of thinking has arisen, is discussed at social events such as group conversations, parties, religious, or town meetings. The uptake of this new idea is driven by social pressure and copying, but because the idea is attractive it will be adopted even if it is held only by a minority of individuals, provided that the minority is not too small.  

The spreading of a new and dominant species (equivalent to a persuasive idea) among a population of mobile agents, driven by a biased majority rule, has previously been studied by Galam \textit{et al.} \cite{Gal98,Gal02,Gal01}. Their focus was on the conditions under which the new species could establish sufficiently large and stable domains to allow expansion of domain walls.  They found that reducing the rate of interaction caused small domains of the new species to dissolve through diffusion, leaving only isolated individuals who died out. In our paper we consider the case where a stable domain is already established and then examine how the velocity of its boundary depends on interaction frequency, bias, and the nature of the random motion of agents. We find that the transmission rate of the new idea, given by the velocity of the domain wall, grows sublinearly---or can even decrease---with increasing frequency of social gatherings. We consider the system in one dimension as a proxy for the motion of a straight domain wall of infinite length in two dimensions (we would expect a curved boundary to introduce surface tension effects \cite{Red10}). 

It has been suggested that the nature of human dispersal leads to scale-free displacements \cite{Bro06}. To capture this our walkers follow truncated symmetric $\alpha$ \textit{stable} processes, also called \emph{L\'{e}vy motion}, or  \emph{L\'{e}vy flights} when considered as a discrete time process \cite{,Con04,Man00,Man94}. The $\alpha$ stable family of processes is parameterized by an index of stability, $\alpha \in (0,2]$, and includes Brownian motion when $\alpha=2$ and anomalous (super) diffusion when $0<\alpha<2$. Our central analytical result is to show that if $\tau$ is the typical time since the last interaction for an arbitrarily selected walker, then the velocity of propagation of the idea obeys $v \propto \tau^{\frac{1}{\alpha}-1}$. When $\alpha<1$, this implies that less-frequent interactions accelerate the dynamics.

The fact that individuals wait for random intervals between interactions has been recognized to have significant effects on the spread of information and ideas through social systems \cite{Kar11,Han14,Gle16}. In particular, bursts of activity \cite{Bar10}, and nonmemoryless waiting times between interactions can alter the rate of spreading, but longer average waiting times slow dynamics \cite{Han14}.  
Our work shows that leaving longer times between interactions does not always slow down dynamics, and we characterize this effect analytically.

The structure of our paper is as follows. In Sec.~\ref{def} we define the processes of social interaction and movement and then present simulation results in Sec.~\ref{sim}. In Sec.~\ref{analysis}, we obtain analytical results for the velocity of propagation of the idea in the limit of large social gatherings. We discuss the implications of our analysis, and give intuitive explanations of our findings in Sec.~\ref{conc}. The appendix provides background and further details on L\'{e}vy and $\alpha$ stable processes.

\section{Model definition}
\label{def}

We consider a population of random walkers, moving in continuous space and time in one dimension. We define $\rho >0$ to be the mean number of walkers per unit length along the line. At all times, each walker is in one of two opinion states $s_k \in \{\up,\dn\}$ where  $k$ indexes the walker. The $\up$ state is dominant and represents acceptance of a new idea which is spreading through the system.  We assume that all walkers are interested in the new idea and it spreads though occasional meetings between groups of nearby walkers. 

\subsection{Gathering process}

We think of these meetings as social gatherings and construct them by introducing a small parameter $\dd >0$ called the \textit{interaction range}. If  $X_k(t)$ is the position of the $k$th walker at time $t$, then we let $\omega(x,t)$ be the set of indices of walkers with locations $X_k(t) \in [x-\dd, x+\dd]$,
\begin{equation}
\omega(x,t) := \{ k \text{ such that } |X_k(t)-x| \leq \dd\}.
\end{equation}
We define a gathering at position $x$ to be a meeting between all walkers in the set $\omega(x,t)$ and note that gatherings of zero size are technically allowed by this definition, but they have no effect on the state of the system. The interaction range serves as a tool for the construction of gatherings and may be thought of as defining what we mean for a group of people to be ``in the same place.'' It should therefore be small compared to the typical distance moved by walkers between interactions. The motion of each walker is assumed to be independent of all others, so at large times their positions form a Poisson point process of intensity $\rho$ along the line. When a gathering starts, the expected group size is therefore 
\begin{equation}
N_G := 2 \rho \delta
\end{equation}
and we control this via the density $\rho$.  In our analytical work we consider the limit of large gatherings.

We assume that gatherings are initiated by a time-space Poisson process which is independent of the trajectories of walkers. That is, the probability that a gathering will take place at some position $x$  within a space interval of length $\dx$ and a time interval $\dt$ is $\lambda \dx \dt$, where $\lambda>0$ is the gathering rate. The expected number of interactions per unit time for each walker is therefore
\begin{equation}
\lambda N_G := \frac{1}{\tau}.
\end{equation}
The quantity $\tau$  is taken as an independent variable in the model, controlling the time scale of interactions, with $\lambda$ dependent on $\tau$. Because the gathering initiation process is independent of the locations of the walkers, then for each walker the waiting times, $T$, between interactions are memoryless and therefore exponentially distributed \cite{Fel67} with density
\begin{equation}
g(t) := \lim_{\dt \ra 0} \frac{1}{\dt}\PP \left\{ T \in (t, t+\dt] \right\} = \frac{ e^{-\frac{t}{\tau}}}{\tau} ,
\label{Tpdf}
\end{equation}
so $\EE[T]= \tau$. To summarize: the two important variables controlling the size and timing of gatherings are $\rho$ and $\tau$. We think of the interaction range just as very small, fixed length scale.   

\begin{figure}
	\begin{center}
		\includegraphics[width=8.5cm]{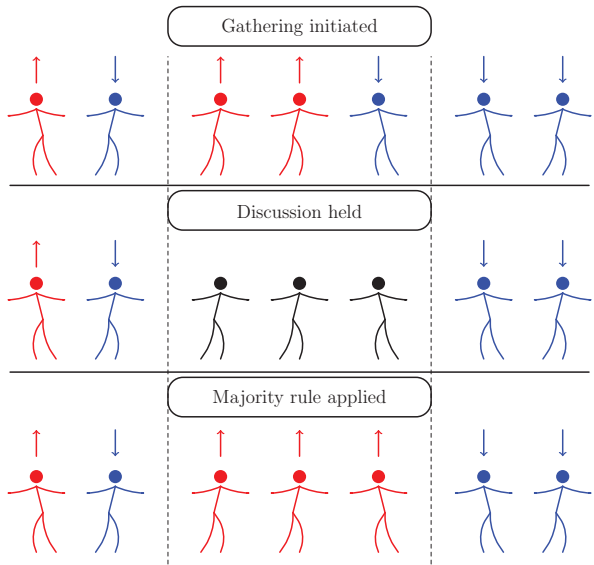}
		\caption{ A symbolic representation of the gathering process. A single gathering is held between the dashed lines, resulting in all walkers accepting the $\up$ opinion. \label{gathering} }
	\end{center}
\end{figure}

After each gathering, all the walkers involved will be of one opinion, decided by a biased version of the majority rule \cite{Kra03}, based on the fraction of walkers in the group who held the new opinion at the start of the meeting. We call this fraction the \textit{opinion field} at position $x$, time $t$,
\begin{equation}
\psi(x,t) := \frac{\sum_{k \in \omega(x,t)} \mathbf{1}_{\{s_k = \up\}}}{|\omega(x,t)|}.
\end{equation}
Here, $\mathbf{1}_{A}$, the indicator of the event $A$, is equal to $1$ if $A$ occurs and zero otherwise; $|\omega(x,t)|$ denotes the number of elements in the set $\omega(x,t)$. Our rule is this: If $\psi(x,t) \ge \tfrac{1-\varepsilon}{2}$, then all walkers in the group adopt the $\up$ opinion, otherwise they all adopt $\dn$. The parameter $\varepsilon \in [0,1]$ represents the strength of the bias toward the dominant opinion. If $\varepsilon>0$, then the $\up$ opinion will tend to spread. Figure \ref{gathering} shows a symbolic representation of the gathering process.

\subsection{Random-walk process}

We consider the case where the paths of walkers follow a L\'{e}vy process \cite{Con04} with L\'{e}vy measure (see the Appendix for more details),
\begin{equation}
\nu(x) = \begin{cases}
c_\alpha|x|^{-(1+\alpha)} &\text{ if } |x| \in (0,l] \\
0 & \text{ otherwise }
\end{cases}
\label{levy_measure}
\end{equation}
where $\alpha \in (0,2)$ and 
\begin{equation}
c_\alpha =
\begin{cases}
\frac{-1}{2 \Gamma(-\alpha) \cos(\alpha \pi/2)} & \text{ if } \alpha \neq 1 \\
\frac{1}{\pi} & \text{ if } \alpha=1.
\end{cases}
\end{equation}
In the limit $l \ra \infty$, this is known as the \emph{symmetric $\alpha$ stable} process. The constant $c_\alpha$ is chosen so the characteristic function of the process in the limit $l \ra \infty$ has a particularly simple form,
\begin{equation}
\EE[e^{i \theta X(t)}] = e^{-t |\theta|^\alpha}.
\label{char}
\end{equation} 
For finite $l$ we will refer to it as \emph{truncated L\'{e}vy motion} (TLM).  The L\'{e}vy measure gives the expected number of jumps with per unit time with sizes $\Delta X \in [a,b]$ as  
\begin{equation}
\int_a^b \nu(x) dx.
\end{equation} 
For TLM the integral of the density $\nu(x)$ over all possible jump sizes does not converge, implying that the expected number of jumps of all sizes in any finite interval is not finite. For this reason the process is  said to have \emph{infinite activity} \cite{Con04}. However, for any $\varepsilon>0$ there are only a finite expected number of jumps having magnitude $|\Dx| > \varepsilon$ in any time interval. The set of jumps for which $|\Dx| < \varepsilon$ form a process which becomes increasingly well approximated by a Brownian motion as $\varepsilon \ra 0$. We may therefore intuitively think of the truncated L\'{e}vy motion as consisting of a diffusion process, plus jumps. 

We define the transition density for TLM using the following notation:
\begin{equation}
L_\alpha^l(x,t) := \PP\{ X(t_0+t) = x_0 + x \mid X(t_0) = x_0\}. 
\end{equation}
When the truncation length $l \ra \infty$ then the transition density for the process has the well known \cite{Man00,Man94} integral representation
\begin{equation}
L_\alpha^\infty(x,t) =\frac{1}{\pi} \int_0^\infty e^{-t q^\alpha} \cos(qx) dq.
\label{levy_pdf}
\end{equation}
The formal relationship among the L\'{e}vy measure, the characteristic function, and the transition probability density for L\'{e}vy processes, and in particular the $\alpha$ stable process, is given in the Appendix. Evaluating the integral in Eq. (\ref{levy_pdf}) with $\alpha=2$, we obtain
\begin{equation}
L_2^\infty(x,t) = \frac{e^{-\frac{x^2}{4 t}}}{2\sqrt{\pi t}},
\end{equation}
so the $\alpha$ stable process with $\alpha=2$ is Brownian motion. 

When the truncation length is finite the properties of the transition density are most easily understood using its characteristic function
\begin{align}
\varphi_t(\theta) & := \EE[e^{i\theta X(t)}] \\
&= \exp \left\{ 2 c_\alpha t \int_0^l (\cos(\theta x)-1) x^{-(1+\alpha)}dx \right\} \\
&= \exp \left\{ 2 c_\alpha t \sum_{n \ge 1} \frac{(-1)^n \theta^{2n} l^{2n-\alpha}}{(2n-\alpha)(2n)!}  \right\}.
\end{align}
From this we see that
\begin{equation}
\EE[X_t] =0 \ \text{ and  } \ \EE[X_t^2] = \frac{2c_\alpha l^{2-\alpha} t}{2-\alpha}.
\end{equation}
On short time scales TLM behaves like the $\alpha$ stable process because the probability of jump sizes $|\Delta X| >l$  in the nontruncated process is small. For large times the central limit theorem implies that the distribution of $X(t)$ becomes progressively more normal with standard deviation,
\begin{equation}
\sigma_0(\alpha,l) \sqrt{t} = \left( \frac{2c_\alpha  }{2-\alpha}\right)^{\frac{1}{2}} l^\frac{2-\alpha}{2} \sqrt{t}.
\end{equation}
Following Ref. \cite{Man94} we equate the probability of return to the origin at time $t$ for a normal (Gaussian) process with standard deviation $\sigma_0(\alpha,l) \sqrt{t}$ to the equivalent probability, $L_\alpha^\infty(0,t)$, for the $\alpha$ stable process,
\begin{equation}
\frac{\Gamma(1+1/\alpha)}{\pi t^{1/\alpha}} = \frac{1}{\sqrt{2 \pi} \sigma_0(\alpha, l) t^{1/2}}
\end{equation}
and solve for $t$
\begin{equation}
t_c = \left( \frac{\sqrt{\pi (\alpha-2) \Gamma(-\alpha)\cos(\pi \alpha/2)} }{\sqrt{2}\Gamma(1+1/\alpha)} \right)^{\frac{2 \alpha}{\alpha-2}} l^\alpha,
\label{crossover}
\end{equation}
where $\alpha \in (0,2)\setminus\{1\}$ and for $\alpha =1$, $$t_c =\frac{4l}{\pi^2} .$$
This gives the approximate crossover time from $\alpha$ stable to normal behavior. The importance of this time will become clear in the remainder of the paper.

\subsection{Opinion wave}

Within our model, large intervals in which all walkers have one opinion are stable against the spontaneous emergence of the other opinion within any subinterval. However, the boundary between two opposing domains will tend to move as the more persuasive opinion domain expands. To investigate the motion of this domain wall, we impose the following initial condition on the opinions of walkers
\begin{equation}
s_k(t_0) = \begin{cases}
\up &\text{ if } X_k(t_0) \leq \Lambda \\
\dn &\text{ if } X_k(t_0) > \Lambda,
\end{cases}
\label{init}
\end{equation}
where $\Lambda \in \mathbb{R}$ gives the starting location of the boundary between the two domains. At $t=t_0$ we allow walkers to start interacting, causing the opinion field to evolve into a smoothed step function. If $\varepsilon>0$, then this step will begin to migrate to the right. Calculating the velocity, $v$, of this propagation is the focus of our work.

\section{Simulation}

\label{sim}

To simulate our model we approximate our TLM with the well-known discrete-time \emph{truncated L\'{e}vy flight} (TLF) process \cite{Man94} which has transition density
\begin{equation}
\PP\{ X_{n+1} - X_n = x\} = \begin{cases}
 c L_\alpha^\infty (x,\Dt) &\text{ if } |x| < l \\
 0 &\text{ otherwise, }
\end{cases}
\end{equation}
where $c$ is a normalizing constant and $\Dt$ may be viewed as the size of the discrete time step. We may approximate the TLM process with a TLF provided we make $\Dt$ sufficiently small so the probability of the TLM increment $X(t+\Dt)-X(t)$ exceeding $l$ is negligible.  

We consider a system of size $L$, chosen so the expected distance traveled by a particle between interactions satisfies $\EE[X(t+T)-X(t)] \ll L$. The system is prepared in state (\ref{init}), where $\Lambda = 0.4 L$ for all simulations, and we allow the simulation to run until the domain wall has reached position $x=0.6L$. In this way, the wall remains distant from the boundaries at all times. We wish to mimic an infinite system where $\lim_{x \ra -\infty} \psi(x,t)=1$, $\lim_{x \ra \infty} \psi(x,t)=0$, and the probability distribution of walker positions is translationally invariant. To achieve this we impose periodic boundary conditions on particle trajectories and maintain the opinion field in opposite states at either end of the system. To maintain the condition on states, when a walker makes a left jump across $x=0$ it is set to state $\dn$, and after a right jump across $x=L$, it is set to $\up$. For example, a walker near the right-hand end of the system, in state $\dn$, on crossing $x=L$ and appearing near the left end of the system, switches to $\up$.

To allow simulations to run within a reasonable time frame we discretize the locations where gatherings can take place to the set $\{2k \dd \}$ where $k \in \mathbb{Z}$, each point being centered on a interval of length $2\dd$ which we refer to as an \emph{urn}. Each urn therefore contains the set of particles which will be involved in a gathering at its midpoint. This allows us to efficiently keep track of which particles are involved in each gathering. At each step of the simulation all particle trajectories are advanced and the urns in which they reside updated. A gathering is then held in each urn with probability $\dt \tau^{-1} \ll 1$, so times between interactions for each walker are well approximated by exponential random variables with mean $\tau$.

\subsection{Zero bias}

\begin{figure}
	\begin{center}
		\includegraphics[width=8.5cm]{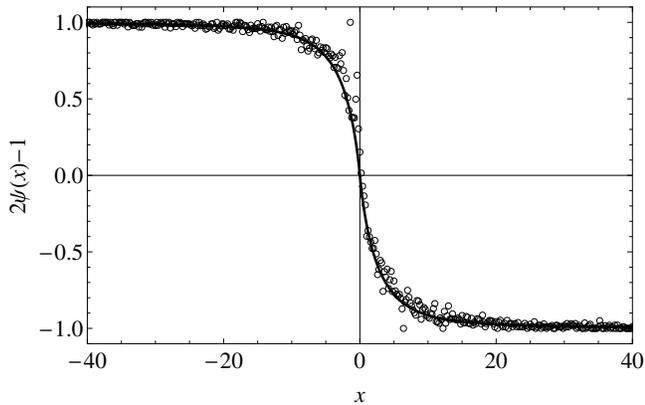}
		\caption{Simulated value of the opinion field when $\alpha=1.5$, $\tau=5$, $\rho = 1000$, and $\delta=0.1$. Plotted points give values of $2 \psi(x,t)-1$ at a grid of $x$ values spaced by $2\delta$. The solid line shows theoretical predicted shape in the limit $\rho \ra \infty$ and $\delta \ra 0$. \label{alpha_15_shape} }
	\end{center}
\end{figure}

\begin{figure}
	\begin{center}
		\includegraphics[width=8.5cm]{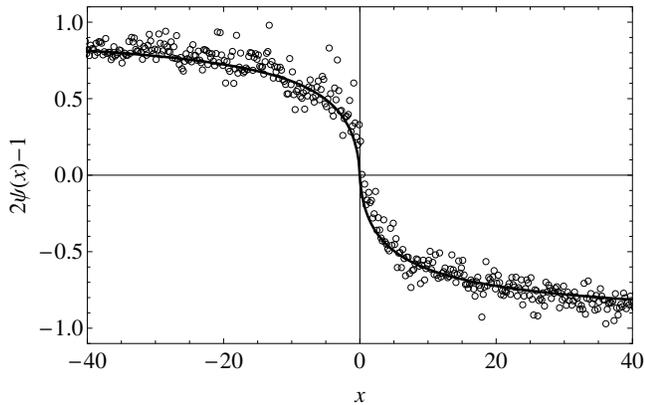}
		\caption{Simulated value of the opinion field when $\alpha=0.75$, $\tau=5$, $\rho = 1000$, and $\delta=0.1$. Plotted points give values of $2 \psi(x,t)-1$ at a grid of $x$ values spaced by $2\delta$. The solid line shows theoretical predicted shape in the limit $\rho \ra \infty$ and $\delta \ra 0$. \label{alpha_075_shape}}
	\end{center}
\end{figure}

We begin by simulating the shape of the opinion field in the case $\varepsilon=0$, which we will see later is crucial to our analytical work on the velocity of propagation. In Fig.~\ref{alpha_15_shape} we have a snapshot of the field values at the centers of each urn for the case $\alpha=1.5$ and $\tau=5$ together with its theoretical shape in the limit $\rho \ra \infty$, derived in Sec.~\ref{analysis}.
In Fig.~\ref{alpha_075_shape} we consider the case $\alpha=0.75$. In both cases we see that, allowing for fluctuations arising from the finite density of walkers, theory and simulation are in close agreement. 

\subsection{Relation among velocity, interaction frequency, and jump distribution}

\begin{figure}
	\begin{center}
		\includegraphics[width=8.5cm]{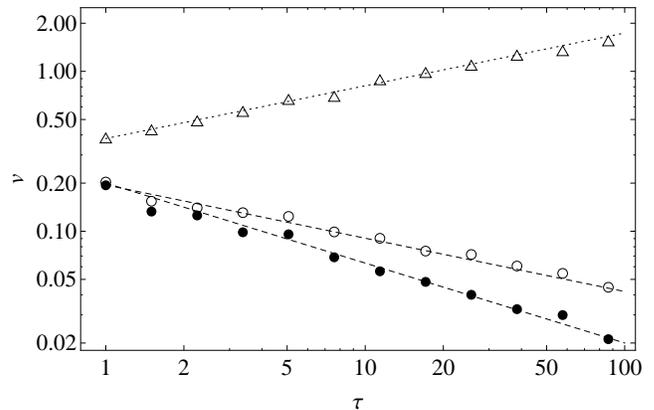}
		\caption{Simulated values of the wave velocity when $\alpha=1.5$ (open circles) and $\alpha=2$ (dots).  In these two cases $\varepsilon=0.1$ and $L=4000$ and dashed lines show full analytical predictions (\ref{velocity}) $v = C \tau^{\alpha^{-1}+1}$ including constant of proportionality. Triangles show wave velocity when $\alpha=0.75$, $\varepsilon=0.2$ and $L=20,000$. Dotted line shows $v \propto \tau^{\alpha^{-1}+1}$ where constant of proportionality is fitted to data. In all cases the particle density is $\rho=40$ and $\delta=0.25$ so each gathering involves $\approx 20$ walkers. \label{alpha_velocity}}
	\end{center}
\end{figure}

We now consider the case $\varepsilon>0$, so the domain wall propagates to the right.   To estimate the velocity of the wall we track its position while it migrates from $x=0.4L$ to $x=0.6L$. We then perform a linear regression on this time series, discarding the early part for which the wall had not yet reached its equilibrium shape. We  explore the behavior of the system when the truncation length, $l$, is sufficiently large so the effects of truncation do no appear while $\tau \in [0,100]$. In Fig.~\ref{alpha_velocity} we have plotted simulation estimates of the propagation velocity for $\alpha \in \{0.75,1.5,2\}$, together with the prediction $v \propto \tau^{\frac{1}{\alpha}-1}$ derived in Sec.~\ref{analysis}. Our results are consistent with the prediction that velocity is an increasing function of interaction rate when $\alpha>1$ but when $\alpha<1$ reducing the frequency of interactions accelerates propagation. 

\subsection{Effect of gathering size}

\begin{figure}
	\begin{center}
		\includegraphics[width=8.5cm]{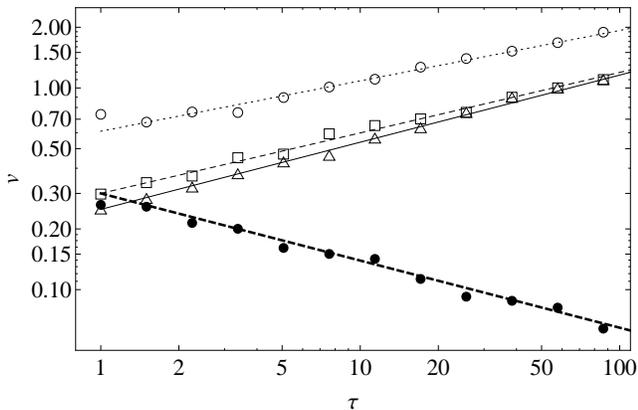}
		\caption{Simulated values of the wave velocity when $\alpha=0.75$ and expected gathering size is $N_G=4$ (open circles, dotted line $v \propto \tau^{0.25}$), $N_G=6$ (open squares, dashed line $v \propto \tau^{0.3}$), and $N_G=8$ (open triangles, solid line $v \propto \tau^{1/3}$). Dots give the $\alpha=1.5$ case when the expected gathering size is $4$, and a thick dashed line gives $v \propto \tau^{-1/3}$.   In all cases $\varepsilon=0.1$ and $\delta=0.25$. \label{small_gather}}
	\end{center}
\end{figure}

Our analytical results of Sec.~\ref{analysis} hold in the limit $\rho \ra \infty$, implying that gatherings occur between infinitely large groups of walkers, leading to a deterministic opinion field. For finite group sizes the opinion field is a discrete valued random variable. We investigate the effects of finite $N_G$ on propagation velocity.  In the simulations of Fig.~\ref{alpha_velocity} we set $N_G= 20$ and found that the relationship $v \propto \tau^{\frac{1}{\alpha}-1}$ remained valid. To investigate the range of values of $N_G$ for which the exponent $\tfrac{1}{\alpha}-1$ in this relationship remains valid we re-run the simulation for smaller gatherings (Fig.~\ref{small_gather}). When $\alpha<1$ we see that values of $N_G<8$ introduce small corrections to the exponent. For $\alpha>1$ the theoretical exponent remains valid down to $N_G=4$. It should be noted that for finite group sizes the velocity does not depend continuously on $\varepsilon$ because of the discrete nature of the opinion field. Such discretization effects become more pronounced for smaller group sizes.

\begin{figure}
	\begin{center}
		\includegraphics[width=8.5cm]{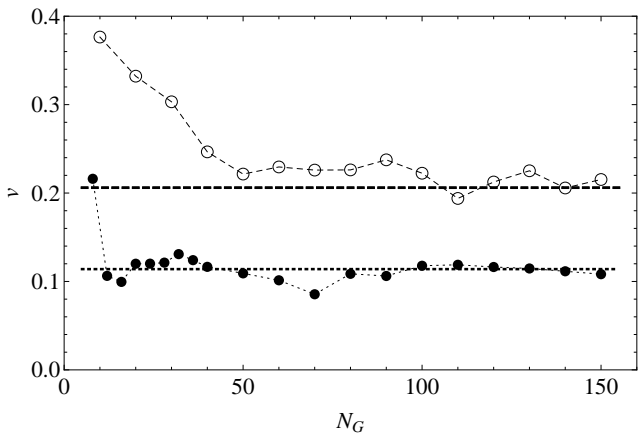}
		\caption{Dependence of wave velocity on gathering size when $\alpha=0.75$ (open circles) and $\alpha=1.5$ (black dots). In both cases $\tau=5$, $\delta=0.25$, system size $L=1000$ and $\varepsilon=0.1$. The thick dotted line shows analytical prediction (\ref{velocity}) and the thick dashed line shows prediction (\ref{vEq}). \label{rho_v}}
	\end{center}
\end{figure}

In Fig.~\ref{rho_v} we have estimated wave velocity for a sequence of $N_G$ values, keeping $\alpha$, $\tau$ and $\varepsilon$ fixed. For the $\alpha>1$ case, we see that our analytical velocity prediction, derived in Sec.~\ref{analysis}, remains valid at least down to gatherings of size $\approx 4$. However, for $\alpha<1$, our analytical prediction overestimates the constant of proportionality between $v$ and $\tau^{\frac{1}{\alpha}-1}$for $N_G \leq 50$.  

The effect of randomized gathering sizes has previously been explored in a nonspatial model of a binary choice public debate \cite{Gal02_2} containing an infinitesimal bias toward the status quo. In this case, varying the distribution of gathering sizes was found to alter the relative concentrations of different opinions at which the two possible final outcomes were equally likely. This effect may be explained by the fact that despite the bias tending to zero, its effect will still be noticed when equal numbers of agents have each opinion, which occurs in even sized gatherings.  In our model, when gathering sizes are small, then we expect this balanced case to occur more frequently, boosting the effect of a small bias.

\subsection{Effects of truncation}

\begin{figure}
	\begin{center}
		\includegraphics[width=8.5cm]{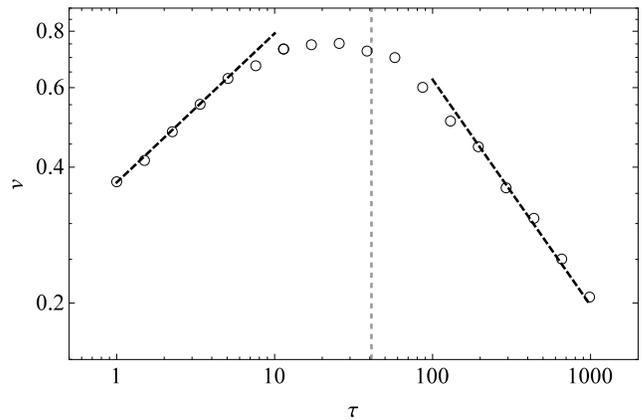}
		\caption{Simulated values of the wave velocity when $\alpha=0.75$ and $l=300$. Particle density is $\rho=40$ and $\delta=0.25$ so each gathering involves $\approx 20$ walkers. System size $L=2 \times 10^4$ and $\varepsilon=0.2$. Heavy dashed lines show analytical predictions $v \propto \tau^{1/3}$ for $\tau \ll t_c$ and $v \propto \tau^{-1/2}$ for $\tau \gg t_c$. Vertical dashed line shows the crossover time $t_c \approx 41$ determined using Eq.~(\ref{crossover}). \label{alpha_075_trunc}}
	\end{center}
\end{figure}

We now reduce the truncation length to the point where the processes begin to appear normal on time scales comparable with $\tau$. In Fig.~\ref{alpha_075_trunc} we have set $\alpha=0.75$ and $l=300$. For low values of $\tau$ propagation velocity exhibits the same $\tau$ dependence as the nontruncated case, but at larger $\tau$ values the velocity exhibits the $\tau$ dependence expected for Brownian motion: $v \propto \tau^{-1/2}$. The crossover occurs because the distribution of TLM switches from $\alpha$ stable to normal at a crossover time $t_c$ given by (\ref{crossover}). We therefore expect a change from a positive to a negative exponent to occur when $\tau = t_c$ and the idea to propagate at maximum velocity at this point.  This prediction is confirmed in Fig.~\ref{alpha_075_trunc}.

\section{Analysis}
\label{analysis}

To compute $\psi(x,t)$ we consider the opinion of walker $k$ at some arbitrary time $t$. Letting $T$ be the (exponentially distributed) time since this walker's last interaction then his or her most recent observation of the opinion field will be
\begin{equation}
\widehat{\psi}_k(t) := \psi(X_k(t-T),t-T),
\end{equation}
and he or she will have the $\up$ opinion if $\widehat{\psi}_k(t)  \ge \tfrac{1-\varepsilon}{2} $, so
\begin{equation}
s_k(t) = \begin{cases}
\up & \text{ if } \widehat{\psi}_k(t)  \ge \frac{1-\varepsilon}{2} \\
\dn & \text{ if } \widehat{\psi}_k(t)  < \frac{1-\varepsilon}{2}.
\end{cases}
\end{equation}
In the limit $\rho \ra \infty$, for any $\delta>0$, the number of walkers in range of any given point, $x$, is infinite, and their historical paths represent an infinitely large sample from the set of all possible paths which reside in the interval $[x-\delta, x+\delta]$ at time $t$. In this case, $\psi(x,t)$ is just the expectation over this set of paths of the fraction of walkers with the $\up$ opinion 
\begin{align}
\psi(x,t) = \PP\left\{ \widehat{\psi}_k(t) \ge \tfrac{1-\varepsilon}{2} \mid X_k(t) \in [x-\delta, x+\delta] \right\}. 
\end{align}
This is now a deterministic quantity. In order to formally take the limit $\delta \ra 0$ while maintaining an infinite number of particles in each gathering, we set
\begin{equation}
\rho = \frac{1}{\delta^2}
\end{equation}
and then consider the behavior of the model as $\delta \ra 0$, in which limit
\begin{equation}
\psi(x,t) = \PP\left\{ \widehat{\psi}_k(t) \ge \tfrac{1-\varepsilon}{2} \mid X_k(t) = x \right\}.
\label{delay}
\end{equation}
We now seek a solution to this equation.

\subsection{Long-time behavior}

Consider the long-term behavior of the system, prepared in initial state (\ref{init}). If $\varepsilon>0$, then the wall will propagate to the right, eventually attaining a constant velocity, $v$, and shape, $f$, so
\begin{equation}
\psi(x,t) \sim f(x-vt) \text{ as } t \ra \infty.
\end{equation}
If we let $X(s)$ be a random walk with $X(0)=0$, and note that $X(s)$ is time reversible, then for large $t$, Eq.~(\ref{delay}) may be written
\begin{equation}
f(x-vt) = \PP\left\{   f(X(T) + x - v(t-T))  \ge \tfrac{1-\varepsilon}{2}  \right\}.
\label{delay2}
\end{equation}	
Making the change of variable $u=x-vt$ in Eq.~(\ref{delay2}) we obtain
\begin{equation}
f(u) = \PP\left\{   f(X(T) + u + v T)  \ge \tfrac{1-\varepsilon}{2}   \right\}.
\label{delay3}
\end{equation}	
Averaging over the probability density function $g$ [see Eq.~(\ref{Tpdf})] of $T$ we may rewrite Eq.~(\ref{delay3}) as
\begin{equation}
f(u) = \int_0^\infty  \PP\left\{   f(X(s) + u + v s)  \ge \tfrac{1-\varepsilon}{2} \right\} g(s) ds.
\label{delay4}
\end{equation}	
Here we have used the following abbreviated notation:
\begin{multline}
\PP\left\{   f(X(T) + u + v T)  \ge \tfrac{1-\varepsilon}{2} \mid T=s\right\}  \\
\equiv \PP\left\{   f(X(s) + u + v s)  \ge \tfrac{1-\varepsilon}{2} \right\}.
\end{multline}
As $\varepsilon \ra 0$, the velocity of propagation tends to zero, so to determine the wave shape in this limit we set $\varepsilon = v = 0$. Without loss of generality we assume that the domain wall was centered on the origin at $t=0$ so $f(0)=\tfrac{1}{2}$ and $f(u) = 1-f(-u)$. Since $f$ is a decreasing function of its argument, then  the condition
\begin{equation}
f[X(s)+u] \ge \frac{1}{2}
\end{equation}
is equivalent to
\begin{equation}
X(s)+u \le 0.
\end{equation}
We define the cumulative of the transition density for our walk 
\begin{equation}
P_s(x) := \PP\{X(t_0+t) < x_0+x | X(t_0)=x_0\},
\end{equation}
and then our delay Eq.~(\ref{delay4}) reads
\begin{align}
f(u) &= \int_0^\infty  \PP\left\{  X(s) + u  \le 0  \right\} g(s) ds \\
&= \int_0^\infty  P_s(-u) g(s) ds. 
\label{shape}
\end{align}
Provided we can evaluate the integral (\ref{shape}), we then have an expression for the wave shape.

Returning now to the case where $\varepsilon >0$ we define
\begin{equation}
Z : = X(s) + u + v s.
\end{equation}
Considering Eq.~(\ref{delay4}), we wish to find the range of values of this variable for which 
\begin{equation}
f(Z) \geq \frac{1-\varepsilon}{2}.
\end{equation}
Since $\varepsilon$ is small and $f(0)=\tfrac{1}{2}$ then the upper bound on $Z$ must be some small positive quantity $\dx > 0$, satisfying $f(\dx) = \tfrac{1-\varepsilon}{2}$. Expanding the opinion wave shape to linear order about the origin, and noting that $f'(0)<0$, we have
\begin{align}
\frac{1}{2} + f'(0) \dx + \mathcal{O}(\dx^2) &= \frac{1-\varepsilon}{2} \\
\dx & \sim \frac{\varepsilon}{2 |f'(0)|} \text{ as } \varepsilon \ra 0.
\end{align}
Making use of this result, delay Eq.~(\ref{delay4}) may now be written 
\begin{equation}
f(u) = \int_0^\infty \PP\left\{   X(s) + u + v s  < \tfrac{\varepsilon}{2|f'(0)|}  \right\} g(s) ds.
\end{equation}
Putting $u=0$ we have
\begin{align}
\frac{1}{2} &= \int_0^\infty \PP\left\{    X(s) + v s - \tfrac{\varepsilon}{2|f'(0)|} < 0  \right\} g(s) ds \\
&= \int_0^\infty P_s\left( \tfrac{\varepsilon}{2|f'(0)|}- v s \right) g(s) ds.
\label{speed}
\end{align}	
After evaluating the integral we can solve for $v$ in terms of $\varepsilon$. We now compute $v$ explicitly.

\subsection{Brownian motion}

When $\alpha=2$ the $\alpha$ stable process reduces to Brownian motion, where increments of the walk are normally distributed at all time scales. This case is of particular importance because any random-walk process with finite variance jump measure has increments whose distribution converges to normal over sufficiently long time steps. Because of its importance and particular tractability we consider this case separately from $\alpha<2$. We also include an additional parameter, $\sigma$, the \emph{volatility} of the walk which is characterized by the transition density
\begin{equation}
p_{\Dt}(\Dx) = \frac{e^{-\frac{\Dx^2}{2 \sigma^2 \Dt}}}{\sqrt{2 \pi \sigma^2 \Dt}}.
\end{equation}
The cumulative of this density is
\begin{equation}
P_{s}(x) = \frac{1}{2} \left( 1 + \erf\left( \frac{x}{\sqrt{2 \sigma^2 s}} \right) \right)
\end{equation}
where the error function is defined
\begin{equation}
\erf(z) := \frac{2}{\sqrt{\pi}} \int_0^z e^{-u^2} du.
\end{equation}
The volatility is related to the diffusion coefficient for a large population of walkers as follows
\begin{equation}
D = \frac{\sigma^2}{2}.
\end{equation}
We note that the $\alpha$ stable process with $\alpha=2$ corresponds to Brownian motion with $D=1$ or, equivalently, $\sigma= \sqrt{2}$. From Eq.~(\ref{shape}) we see that the shape of the traveling wave in the limit $\varepsilon \ra 0$ is
\begin{equation}
f(u) = \begin{cases}
\frac{1}{2} \exp \left\{ -\sqrt{\frac{2}{\tau}}\frac{u}{\sigma}\right\} & \text{ if } u>0 \\
1 - \frac{1}{2} \exp \left\{ \sqrt{\frac{2}{\tau}}\frac{u}{\sigma}\right\}  & \text{ if } u \leq 0,
\end{cases}
\end{equation} 
so the magnitude of the derivative of the wave shape at the origin is
\begin{equation}
|f'(0)| = \frac{1}{\sigma \sqrt{2 \tau}}.
\end{equation}
We now compute the wave speed by evaluating the integral in Eq.~(\ref{speed}) to give
\begin{equation}
\exp \left\{\varepsilon \sigma \sqrt{\frac{\tau}{2}} \left(v- \sqrt{v^2 + \frac{2}{\tau}} \right) \right\} \left(1 + v \sqrt{\frac{\tau}{2 + v^2 \tau}} \right) =1.
\end{equation}
We wish to solve this equation form $v$ when $\varepsilon$ is small. We therefore Taylor expand the left-hand side to linear order about $\varepsilon=0$ and $v=0$ and then solve for $v$, giving
\begin{equation}
v = \varepsilon \sigma \sqrt{\frac{2}{\tau}} = 2 \varepsilon \sqrt{\frac{D}{\tau}}.
\label{v_brown}
\end{equation}
From this, we see that when walkers undergo Brownian motion, then greater levels of diffusion accelerate the wave, and that longer waits between interactions slow it down. See Fig.~\ref{alpha_velocity} for confirmation of this result.

\subsection{L\'{e}vy motion for $l \ra \infty$}

We consider the case of nontruncated jump measure ($l \ra \infty$). In this case, the transition density is $L_\alpha^\infty(x,t)$ [Eq.~(\ref{levy_pdf})], with corresponding cumulative
\begin{equation}
P_t(x) := \frac{1}{2} + \frac{1}{\pi} \int_0^\infty e^{-t q^\alpha} \frac{\sin(qx)}{q} dq.
\end{equation}
Equation (\ref{shape}) then gives the shape of the traveling wave
\begin{equation}
f(u) = \frac{1}{2} - \frac{1}{\pi} \int_0^\infty \frac{\sin(q u)}{q(1+q^\alpha \tau)} dq,
\label{as_wave}
\end{equation}
from which we can extract its derivative at the origin,
\begin{equation}
|f'(0)| = \frac{\csc(\pi/\alpha)}{\alpha \tau^{\frac{1}{\alpha}}} \text{ for } 1< \alpha <2.
\end{equation}
We note that $|f'(0)| \ra \infty$ as $\alpha \ra 1^+$  so our earlier analysis of wave velocity breaks down in this limit. For $\alpha<1$ the wave shape has an infinite derivative at the origin and we adapt our analysis to deal with this case separately. See Fig.~\ref{alpha_15_shape} and \ref{alpha_075_shape} for verification of the wave shape function (\ref{as_wave}) by simulation.

\subsubsection{$\alpha>1$ case.}

To compute the wave velocity, we set $\kappa = \varepsilon/(2 |f'(0)|)$  and evaluate Eq.~(\ref{speed}),
\begin{align}
\frac{1}{2} &= \int_0^\infty P_s\left( \kappa- v s \right) g(s) ds \\
&= \frac{1}{2} + \frac{1}{\pi} \int_{0}^{\infty} \frac{dq}{q} \left[ \int_0^\infty e^{-s q^\alpha} \sin(q(\kappa - v s)) g(s) ds \right] \\
&=\frac{1}{2} + \frac{1}{\pi} \int_{0}^{\infty}  
\frac{\left(\tau  q^{\alpha }+1\right) \sin (\kappa  q)-q \tau  v \cos(\kappa  q)}
{q^2 \left(\tau  \left(\tau  q^{2 \alpha }+2 q^{\alpha }+q^2 \tau  v^2\right)+1\right)} dq.
\label{intract}
\end{align}
The integral (\ref{intract}) over $q$ is intractable, but we can exploit the fact that $\kappa \ra 0$ as $\varepsilon \ra 0$ and expand the integrand as a Taylor series in these two variables, to first order about $\kappa=0$, $v=0$. After integration, we obtain
\begin{equation}
\frac{1}{2} \sim \frac{1}{2} + \frac{\pi  \tau ^{-1/\alpha } \csc \left(\frac{\pi }{\alpha }\right)
	(\alpha  (\kappa -\tau  v)+\tau  v)}{\alpha ^2} \text{ as } \varepsilon \ra 0.
\end{equation}
Solving for $v$, and replacing $\kappa$ with its definition in terms of $\varepsilon$ yields
\begin{equation}
v  \sim \frac{\alpha ^2 \varepsilon  \tau ^{\frac{1}{\alpha }-1} \sin \left(\frac{\pi}{\alpha }\right)}{2 (\alpha -1)} \text{ as } \varepsilon \ra 0.
\label{velocity}
\end{equation}
See Fig.~\ref{alpha_velocity} for confirmation of this result by simulation. Provided $\alpha>1$ then the wave velocity is a decreasing function of the expected time between interactions. Making use of the limit
\begin{equation}
\lim_{\alpha \ra 2} \frac{\alpha ^2  \sin \left(\frac{\pi}{\alpha }\right)}{2 (\alpha -1)} = 2
\end{equation}
we see that as $\alpha \ra 2$ then $v \sim 2 \varepsilon \tau^{-\frac{1}{2}}$ consistent with our earlier result for Brownian motion $(\ref{v_brown})$ when $D=1$.

\subsubsection{$\alpha \leq 1$ case}

When $\alpha \leq 1$, then the singular behavior of $f'(u)$ at the origin forces us to reconsider our solution method for 
\begin{equation}
f(Z) = \frac{1-\varepsilon}{2}.
\label{Zeqn}
\end{equation}
An alternative avenue is to make use of the scaling properties of the wave shape function. We begin by defining the wave shape when $\tau=1$,
\begin{equation}
f_1(u) := \frac{1}{2} - \frac{1}{\pi} \int_0^\infty \frac{\sin(q u)}{q(1+q^\alpha)} dq.
\end{equation}
We now note that
\begin{align}
f_1(\tau^{-\frac{1}{\alpha}}u) &= \frac{1}{2} - \frac{1}{\pi} \int_0^\infty \frac{\sin(q \tau^{-\frac{1}{\alpha}}u)}{q(1+q^\alpha)} dq  \\
&= \frac{1}{2} - \frac{1}{\pi} \int_0^\infty \frac{\sin(y u)}{y(1+ y^\alpha \tau)} dy \\
& = f(u),
\end{align}
where we made the change of variable $y = \tau^{-\frac{1}{\alpha}} q$. From this scaling relationship we see that Eq.~(\ref{Zeqn}) has the solution
\begin{equation}
Z = \tau^{\frac{1}{\alpha}}f_1^{-1}\left(\frac{1-\varepsilon}{2}\right):= C(\alpha,\varepsilon)  \tau^{\frac{1}{\alpha}},
\end{equation}
where $f_1^{-1}$ is the inverse function of $f_1$ and $C(\alpha,\varepsilon)>0$ is independent of $\tau$. To complete our analysis, we must solve for $v$ in the equivalent of Eq.~(\ref{speed}),
\begin{equation}
\frac{1}{2}  = \int_0^\infty P_s\left( C(\alpha, \varepsilon) \tau^{\frac{1}{\alpha}}- v s \right) g(s) ds.
\end{equation}
We begin by noting that the cumulative of the L\'{e}vy density satisfies the same scaling relationship as the wave shape, that is, 
\begin{equation}
P_s(x) = P_1(s^{-\frac{1}{\alpha}} x).
\label{Pscale}
\end{equation}
We now make the ansatz that
\begin{equation}
v = A(\alpha, \varepsilon) \tau^{\frac{1}{\alpha}-1},
\label{vEq}
\end{equation}
where $A(\alpha, \varepsilon)$ is constant, to be determined. Making use of this ansatz, and the scaling property (\ref{Pscale}), Eq.~(\ref{vEq}) becomes
\begin{align}
\frac{1}{2} &= \int_0^\infty P_1\left[C(\alpha, \varepsilon) \left(\frac{\tau}{s}\right)^{\frac{1}{\alpha}} - A(\alpha, \varepsilon) \left(\frac{\tau}{s}\right)^{\frac{1}{\alpha}-1} \right] \frac{e^{-\frac{s}{\tau}}}{\tau} ds \\
&= \int_0^\infty P_1\left[ C(\alpha, \varepsilon) y^{-\frac{1}{\alpha}} - A(\alpha, \varepsilon) y^{1-\frac{1}{\alpha}} \right] e^{-y} dy.
\label{AEq}
\end{align} 
Since $P_1$ is an increasing function with $P_1(0)=\tfrac{1}{2}$, then there must be an $A(\alpha, \varepsilon)>0$ for which (\ref{AEq}) holds, demonstrating that our ansatz was correct. The constant of proportionality  $A(\alpha,\varepsilon)$ may be found by numerical solving Eq.~(\ref{AEq}) and is verified by simulation for large $\rho$ in Fig.~\ref{rho_v}.  

We have now established the general relationship for nontruncated L\'{e}vy motion,
\begin{equation}
v \propto \tau^{\frac{1}{\alpha}-1}
\end{equation}
where $\alpha \in (0,2]$. This result is verified by simulation in Fig.~\ref{alpha_velocity} and leads to the counter-intuitive result that reducing interaction rate can accelerate the opinion wave, provided $\alpha<1$.

\subsection{Truncated L\'{e}vy motion}

In any finite system, jump sizes are necessarily truncated. Because large jumps are very rare, over short time scales the TLM behaves statistically like the nontruncated process with a crossover to normal transition probabilities at time $t_c$ given by Eq.~(\ref{crossover}). Therefore we expect to see a transition from L\'{e}vy to normal velocity behavior when the time between gatherings is sufficiently long; that is,
\begin{equation}
v \propto \begin{cases}
\tau^{\frac{1}{\alpha}-1} &\text { when } \tau \ll K l^\alpha \\
\tau^{-\frac{1}{2}} &\text { when } \tau \gg K l^\alpha \\
\end{cases},
\end{equation}
where for $\alpha \neq 1$
\begin{equation}
K = \left( \frac{\sqrt{\pi (\alpha-2) \Gamma(-\alpha)\cos(\pi \alpha/2)} }{\sqrt{2}\Gamma(1+1/\alpha)} \right)^{\frac{2 \alpha}{\alpha-2}},
\end{equation}
and $K = 4/\pi^2 $ when $\alpha=1$. When $\alpha<1$ this leads to a peak velocity at critical interaction frequency. This effect is illustrated in Fig.~\ref{alpha_075_trunc}.

\section{Discussion and Conclusion}
\label{conc}

We have considered a simple model of the spread of an idea through a population of mobile individuals with scale-free displacements truncated at size $l$. The idea is exchanged at social gatherings of nearby individuals using a biased majority rule. We have derived two main results in the limit of large gatherings:
\begin{enumerate}
	\item
	The velocity of propagation obeys $v \propto \tau^{\frac{1}{\alpha}-1}$, where $\tau $ is the expected time between interactions for a single walker and $\alpha \in (0,2]$ characterizes the random motion. 
	\item 
	For finite $l$, the velocity exhibits a crossover from L\'{e}vy to normal velocity behavior at critical interaction rate.
\end{enumerate}
These results have been confirmed by simulations, which also show that they remain valid down to quite modest gathering sizes.

In summary, the model predicts that if movement of walkers is merely diffusive, \emph{i.e.}, gatherings typically involve groups of walkers who have not traveled far since their last interaction, then persuasive ideas that are frequently discussed will spread faster than those discussed less frequently. This is a result that might have been expected. However, the model also predicts that when movement of walkers follows a L\'{e}vy process that is sufficiently \textit{superdiffusive}, or ``jumpy,'' then the opposite is true: Infrequently discussed persuasive ideas spread faster than those discussed more often (provided $\alpha<1$). Furthermore, the more ``jumpy'' the movement of walkers, the greater this effect, unless jumps are truncated at some maximum size; then we find there is an optimal frequency of discussion. In addition, we also find that the idea will spread faster when discussions are held in smaller groups. To put it in another way, according to our model, the ideas that spread fastest are those discussed infrequently, by small groups of walkers, whose movement follows a particularly ``jumpy'' L\'{e}vy process. 


We offer the following explanation of the effect: Using past discussions to decide current behavior results in a propagation velocity for new ideas which is determined by two competing factors: infrequent discussion anchors current behavior to the past, slowing down the rate of change, but it also allows individual opinions from far afield to be spread long distances. In fact, we may see how these two competing effects combine to produce our main result using simple dimensional analysis. We have two important quantities on which the domain wall velocity depends: The typical time, $\tau$, between interactions, and the typical distance $d(\tau)$ that each walker travel between interactions. From the scaling properties of the L\'{e}vy transition density (\ref{levy_pdf})
\begin{equation}
L_\alpha^\infty(x,t) = L_\alpha^\infty(t^{-\frac{1}{\alpha}} x,1) 
\end{equation}
we have that $d(\tau) \propto \tau^{\frac{1}{\alpha}}$. Combining our distance and time variables to produce the dimensions of velocity, we have
\begin{equation}
v \propto \frac{d(\tau)}{\tau} \propto \tau^{\frac{1}{\alpha}-1}.
\end{equation}
From this we see that increasing the time between interactions contributes in opposing ways to the velocity. It increases the distance the idea can travel between discussions but reduces the rate of discussion.  

If our model reflects reality, then we would expect to find that frequently discussed ideas have a narrower range of possible spreading rates than those ideas discussed less often. Less frequently discussed ideas are more likely to be either very fast or very slow to spread. To test this, one might look for data on the spread of ideas, opinions, dialect and language features, or other behaviors, and in particular look at whether any correlation exists between the rate of spread and the frequency of discussion. The nature of that correlation, or even its absence altogether, could provide insight into the process that might be at work. A potential example of such an effect is the outward spread of new words or other language features from an economically and culturally dominant center \cite{Bri03}. In this case, we would expect the frequency with which the word was used, as well as the time required to learn it, to play a role in how quickly it spread.

The simplicity of our model allowed the derivation of analytical results, which invite further investigation for more realistic models. For example, it may be that individuals base their current behavior on a whole series of historical social interactions, rather than on just the latest one or that waiting times between social interactions may not be memoryless, associated with bursts of social activity \cite{Iri09,Bar10,Lam13}.

\section*{ACKNOWLEDGMENTS}
J.B. is grateful for the support of a Leverhulme Trust
research fellowship.
\appendix*

\section{FULL DEFINITION OF SYMMETRIC $\alpha$ STABLE L\'{E}VY MOTION}	
\label{LevyDef}

Here we provide a brief but complete definition of symmetric $\alpha$ stable L\'{e}vy motion and the connection between its L\'{e}vy measure and transition probability density. For more details on  L\'{e}vy and stable processes we refer the reader to Refs.~\cite{App04} and \cite{Sam94}.

In general, a \emph{L\'evy process} $X=(X(t))_{t \geq 0}$ is a stochastic process with initial value $X(0)=0$ and the following properties:
\begin{enumerate}
	
	\item 
	\textbf{Independent increments.} For any sequence of times $0 \leq t_0<t_1<t_2 < \ldots < t_n < \infty$ the increments $X(t_1) - X(t_0), X(t_2) - X(t_1) , \ldots , X(t_n) - X(t_{n-1}) $ are independent.
	
	\item 
	\textbf{Stationary increments.} Let $t\geq 0$ and $s \geq 0$. The distribution of $X(t+s)-X(t)$ does not depend on $t$. 
	
	\item 
	\textbf{Stochastic continuity.} For any $\varepsilon>0$, $ \lim_{h \dn 0} \PP(|X(t+h)-X(t)|>\varepsilon) = 0$. 
	
\end{enumerate} 

The transition probability density, $p_h(x)$, for a L\'{e}vy process is defined as
\begin{equation*}
p_h(x) =  \lim_{\dx \ra 0} \frac{1}{\dx} \PP\left\{X(t+h)-X(t) \in [x, x+\dx]\right\} 
\end{equation*}
and can be derived as the inverse Fourier transform of the complex conjugate of the characteristic function, $\varphi_t(\theta)$, of the process
\begin{align*}
\varphi_t(\theta) &= \mathbb{E}\left[e^{i\theta X(t)}\right]\\ 
p_h(x) &= \frac{1}{2\pi} \int_{-\infty}^{\infty} e^{i \theta x}\overline{\varphi_t(\theta)} d\theta \\
&= \frac{1}{2\pi} \int_{-\infty}^{\infty} e^{-i \theta x}\varphi_t(\theta) d\theta .
\end{align*}
The connection between the L\'{e}vy measure, $\nu$, and the characteristic function $\varphi_t$ of the process is given by the \emph{L\'evy--Khintchine formula},  as follows:
\begin{equation*}
\varphi_t(\theta) =\exp{\left\{t\left(bi\theta -\frac{1}{2}\sigma^2\theta^2+\int_{\mathbb{R}\setminus \{0\}} g(x, \theta) \nu(dx)   \right)\right\} }, 
\end{equation*}
where $g(x, \theta)  =e^{i\theta x}-1-i\theta x \mathbf{1}_{|x|<1} $, $b\in \mathbb{R}$ is a location parameter, $\sigma^2$ is a Gaussian variance, and $\nu$ is a  measure such that 
\begin{equation*}
\int_{\mathbb{R}\setminus \{0\}} \min\left\{|x|^2,1\right\} \nu(dx) <+\infty.
\end{equation*}
We refer to the triple $(b, \sigma, \nu)$ as the \emph{characteristics of a L\'evy process} $X$. In the case when
\begin{equation*}	
\int \nu(dx) = \int f(x) dx,
\end{equation*}
that is, the measure $\nu$ is induced by a density $f$ with respect to the Lebesgue measure $dx$, we denote this density also by $\nu$ and refer to it as a L\'evy measure.

A random variable $Y$ is said to be \emph{strictly stable}  if, for any  $a,b>0$, there exists a number $c>0$ such that $aY_1 + b Y_2$ has the same distribution as $cY$, where $Y_1$ and $Y_2$ are independent copies of $Y$. In particular, there exists $\alpha \in (0,2]$ such that $a^{\alpha}+b^{\alpha} = c^{\alpha}.$ 
The number $\alpha$ is called \emph{the index of stability}. If $Y$ and $-Y$ have the same distribution, then we refer to $Y$ as a \emph{symmetric stable} random variable. 

We say that a L\'evy process is a \emph{symmetric stable L\'evy motion} if each $X(t)$ is a symmetric stable random variable. In particular, we classify symmetric stable processes according to the index of stability $\alpha \in (0,2]$ of $X(1)$. Namely, we associate $\alpha$ with $X$ if $t^{-1/\alpha}X(t)$ has the same distribution as $X(1)$ for each $t$, and call $X$ a \emph{symmetric $\alpha$ stable process}. 
For $\alpha \in (0,2)$ the characteristics of $X$ is $(0, 0, \nu)$ with  the  L\'evy measure defined as 
\begin{equation}
\label{levymeasure0}
\nu(x)= c |x|^{-(1+\alpha)}
\end{equation}
for $x \in \mathbb{R}\setminus\{0\}$, where $c>0$. For $\alpha = 2$ the characteristics of $X$ is $(0,\sigma,0)$ and in particular if $\sigma = 1$, then for each $t$, $X(t)$ has normal distribution $\mathcal{N}(0,t)$ and so $X$ is a standard Brownian motion. 

Applying the L\'evy--Khintchine formula with the  L\'evy measure (\ref{levymeasure0}) we obtain the following characteristic function of $X(t)$:
\begin{equation}\label{character}
\varphi_t(\theta; \alpha,s) = \exp(- s^{\alpha}|\theta|^{\alpha}t),
\end{equation}
where $s>0$ is a scale parameter depending on a constant $c$ from (\ref{levymeasure0}). 
For simplicity we consider a symmetric $\alpha$-stable L\'evy motion with a characteristic function $\varphi_t(\theta; \alpha, 1)$. To obtain that, that is, to make the scaling parameter $s$ to equal $1$, we take the L\'evy measure of the form 
\begin{equation}
\nu(x)=c_{\alpha}|x|^{-(1+\alpha)},
\end{equation}
where
$$c_{\alpha} = \begin{cases} -\frac{1}{2\Gamma(-\alpha)\cos\left(\frac{\pi \alpha}{2} \right)} & \mbox{ for } \alpha \neq 1\\ 
\frac{1}{\pi} & \mbox{for } \alpha = 1 
\end{cases}. $$
The formula for $c_{\alpha}$ can be derived by applying the L\'evy--Khintchine formula and using the fact that
$$ \int_0^{\infty} (\cos(\theta x) -1 ) x^{-(1+\alpha)} dx = |\theta|^{\alpha}\Gamma(-\alpha)\cos\left(\frac{\pi \alpha}{2} \right)$$
for $\alpha \in (0,2 )\setminus\{1\}$.

\bibliography{opinion_diff_refs}

\end{document}